\documentclass[%
 reprint,
 amsmath,amssymb,
 aps,
 prl,
 secnumarabic,
]{revtex4-2}

\usepackage[utf8]{inputenc}
\usepackage{graphicx}
\usepackage{subcaption}
\usepackage{dcolumn}
\usepackage{bm}
\usepackage{comment}

\begin{document}

\preprint{APS/123-QED}

\title{Piecemeal method revisited}

\author{Yi-Xin Shen}
\affiliation{State Key Laboratory of Low Dimensional Quantum Physics, Department of Physics, Tsinghua University, Beijing 100084, China}

\author{Zhou-Kai Cao}
\affiliation{State Key Laboratory of Low Dimensional Quantum Physics, Department of Physics, Tsinghua University, Beijing 100084, China}

\author{Jian Leng}%
\affiliation{State Key Laboratory of Low Dimensional Quantum Physics, Department of Physics, Tsinghua University, Beijing 100084, China}

\author{Xiang-Bin Wang}
\email{xbwang@mail.tsinghua.edu.cn}
\affiliation{State Key Laboratory of Low Dimensional Quantum Physics, Department of Physics, Tsinghua University, Beijing 100084, China}


\date{\today}

\begin{abstract}
Detecting the angles and orbits of remote targets precisely has been playing crucial roles in astrophysical research.
Due to the resolution limitations imposed by the Airy disk in a single telescope, optical interferometric schemes with at least two telescopes have received considerable attention.
We have extended the piecemeal method to reduce the required number of baselines for observation. Through the analysis of its performance under practical conditions, we demonstrate that both the original and extended piecemeal methods exhibit strong robustness against errors in baseline lengths and orientations. Under the same practical conditions, our approach achieves higher precision than other existing weak-light interference-based methods.
\end{abstract}

\maketitle

\section{Introduction}
Observation of the orbital diameter, orbital period, and orbital precession of remote targets as quickly and accurately as possible is an important task in cosmology research. The observation precision of a single telescope is limited by the Rayleigh criterion: $R=\frac{\lambda}{D}$, where $\lambda$ is the wavelength of the received light, and $D$ is the diameter of the telescope. Evidently, to get higher angular resolution and observe stars with higher magnitudes, we have to increase the aperture of the telescopes. However this cannot be done unlimitedly. Methods based on optical interference have drawn much attention. Nowadays, large-baseline optical interferometers around visible light band, such as VLTI\cite{glindemann2001light, le2009first}, 
Keck\cite{colavita2003observations}, CHARA\cite{ten2005first}, and GRAVITY\cite{abuter2017first} are built and used by numerous institutes across the world. These interferometers measure the visibility of interference fringes and use \textit{van-Cittert/Zernike theorem} to invert the distribution of light sources in space \cite{goodman_statistical_nodate}. Currently, their baseline lengths can be hundreds of meters; however, the presence of noise and transmission loss limits the length of their baseline to increase further. Therefore, employing weak-light-interference measurement techniques, which require fewer photons, would be a significant improvement for existing telescopes.
In recent years, a variety of schemes based on single-photon/weak-light interference have been proposed, including \cite{gottesman_longer-baseline_2012, khabiboulline_optical_2019, huang_imaging_2022, marchese_large_2023}.
However, the observation precision of all these single-baseline methods is limited by various imperfections in practice. There are numerous uncontrollable errors for a real setup, for example, variations in the baseline length and orientation of the telescope, as well as changes in optical path differences caused by fluctuations in atmospheric density and thickness. Under these noises, the relative phases of the photons collected by the telescope are usually different from the expected value and cannot provide high-precision results.

Differently, our piecemeal method \cite{leng_piecemeal_2024} can reach a higher-level precision because it uses observation data from different baselines jointly.
In this work, we study our piecemeal protocol with reference star under practical set-up imperfections, including errors in baseline lengths and orientations. With the help of reference stars, our protocol has better resolution than existing single-baseline protocols under the influence of practical factors.

The structure of this paper is as follows: In Section 2, we generalize our piecemeal method with reference stars to allow the length ratio between the two adjacent baselines to be any integer multiple. Sections 3 focus on the impact of set-up imperfections, demonstrating the robustness of our scheme against baseline length and orientation errors. Finally, in Section 4, we summarize the advantages of our current scheme and outline potential directions for future improvements.

\section{Extended piecemeal method}\label{sec:piecemeal_description}

\subsection{Generalization of piecemeal method}
The piecemeal method introduced in \cite{leng_piecemeal_2024} is an interferometric approach that enables precise angle measurement using a relatively small number of photons by incorporating multiple baselines of different lengths. 
We extend this method to a more general case: The length ratio between two adjacent baselines is no longer restricted to 2 but can be any integer multiple. The generalized method reduces the dependence on the number of baselines.

\begin{figure}[ht]
    \centering
    \begin{minipage}[t]{0.9\linewidth}
        \centering
        \includegraphics[width=\linewidth]{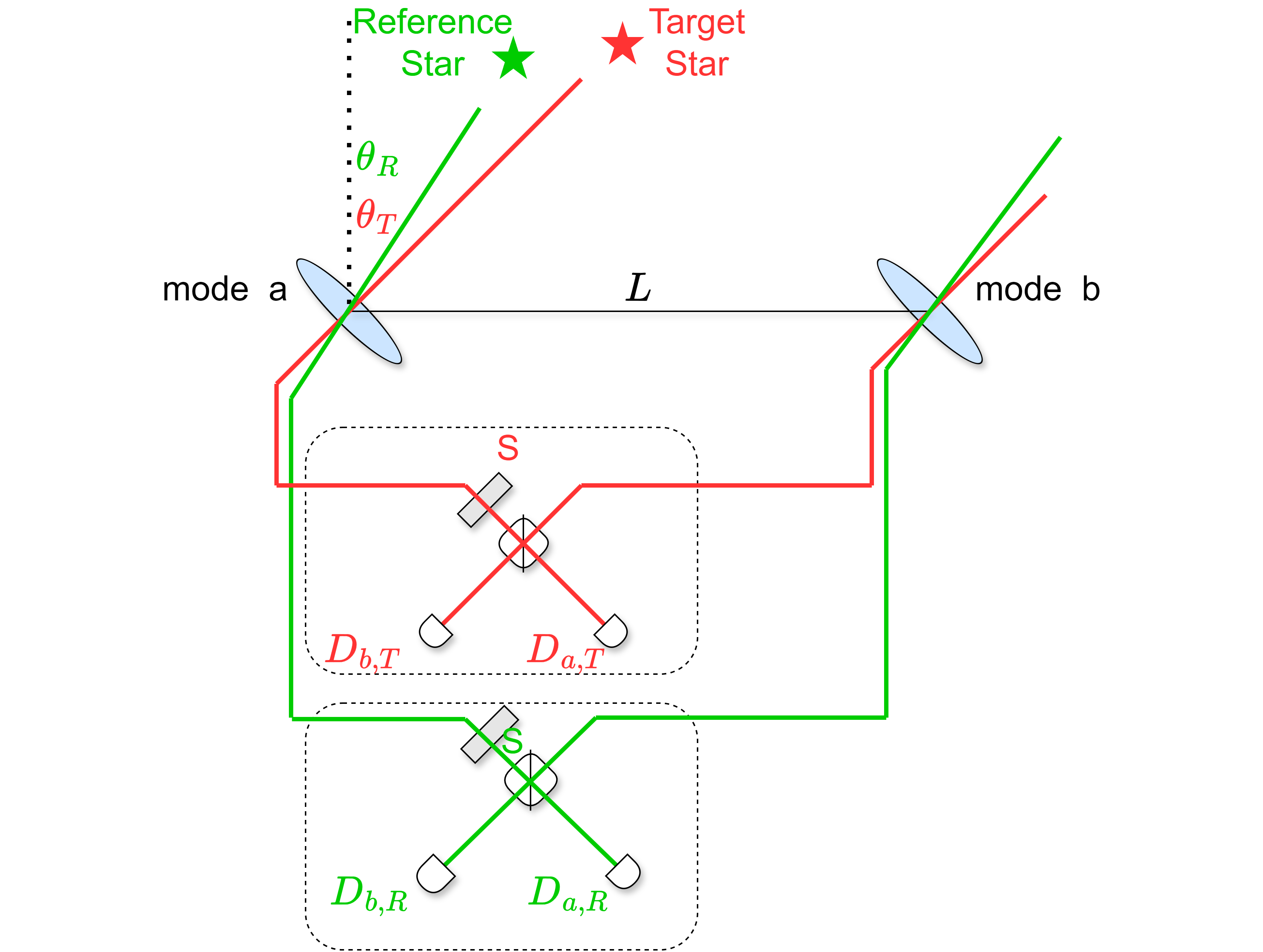}
        \caption*{(a)}
    \end{minipage}
    \begin{minipage}[t]{0.75\linewidth}
        \centering
        \includegraphics[width=\linewidth]{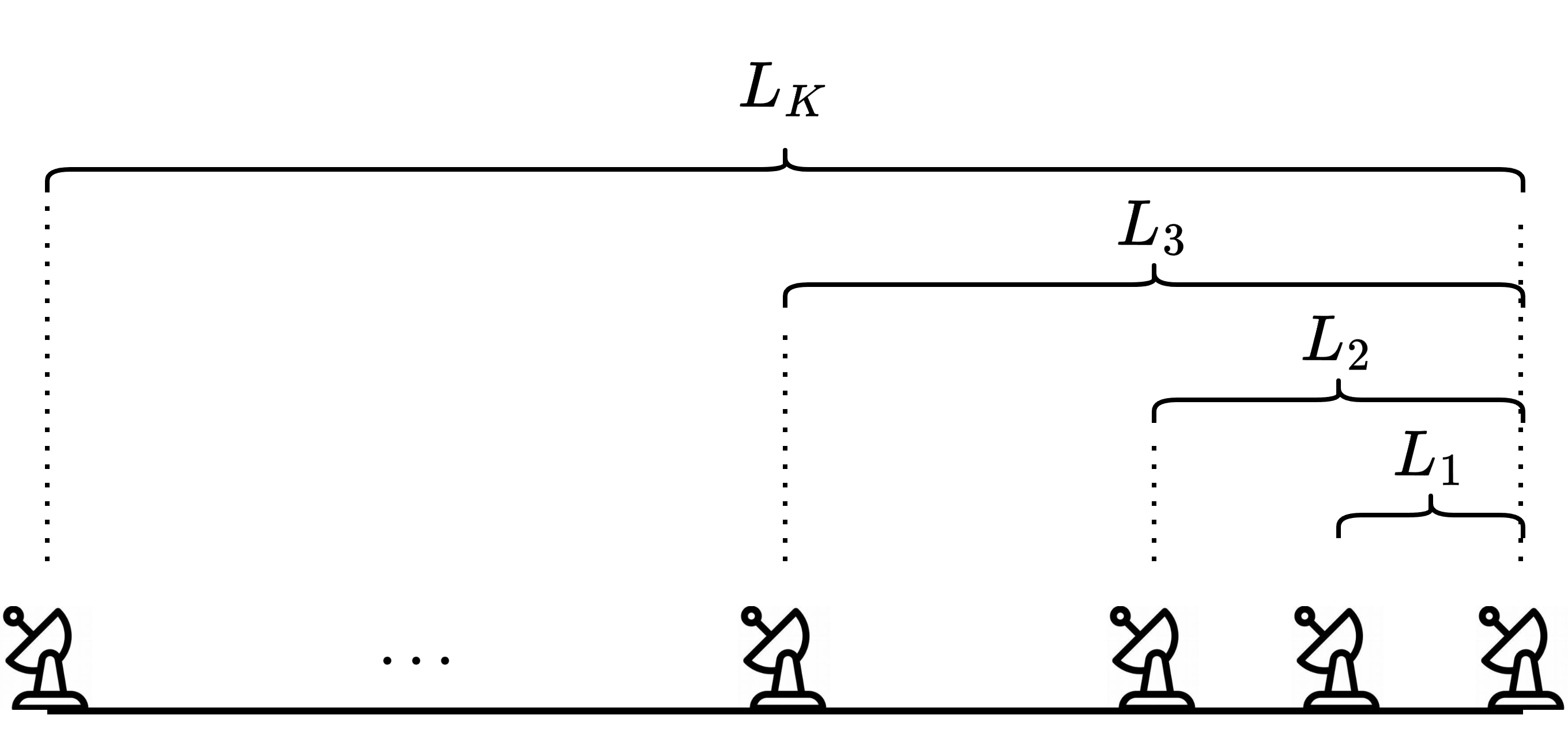}
        \caption*{(b)}
    \end{minipage}
    \caption{Schematic setup of our protocol that utilizing a reference star to assist in eliminating channel noise. (a) Photons from the target star and the reference star enter different beam splitters and are respectively measured by two threshold detectors at the output port of each beam splitter. (b) Our protocol contains different baselines $\{L_k\}$. The lengths of two adjacent baselines are designed to differ by integer multiples: $L_k=s_k L_{k-1}$. Data from different baselines are collected separately but processed jointly.} 
    \label{fig:SingleLayerSetup}
\end{figure}

The set-up of our piecemeal method is schematically shown in Fig.\ref{fig:SingleLayerSetup}. Our method consists of \(K\) baselines, each with a telescope positioned at both ends. We take data from each baseline independently and process them jointly. 
The length of the $k-th$ baseline is $L_k$, where $k=\{1,2\cdots K\}$. The lengths of two adjacent baselines are integer multiples of each other: \( L_{k+1} = s_k L_k \).

Suppose that the angle of remote object is $\theta$. Suppose the weak-light state collected by the pair of telescopes at both ends of baseline $k$ attenuates to the single-photon state
\begin{equation}\label{equ:Sk_ideal}
    |\mathcal{S}_k\rangle=\frac{1}{\sqrt{2}}(|01\rangle+\mathrm{e}^{i\phi_k}|10\rangle),
\end{equation}
where $\phi_k=2\pi \frac{L_k}{\lambda} \theta$, $\lambda$ is the wavelength of the incident single photon, and $\left|c_a c_b\right\rangle$ represents photon numbers of mode $a$ and $b$.

\textbf{Notations}:
\begin{enumerate}
    \item Taking $L_{k+1}=s_k L_{k}$ and hence $\phi_{k+1}=s_k \phi_k$.
    \item $\mathring{x}=x \bmod 2 \pi$. When using $\alpha \mathring{x}$, we mean $\alpha(x \bmod 2 \pi)$.
    \item The observed value of $\mathring{x}_k$ is denoted as $\hat{x}_k$.
\end{enumerate}

In the observation of each baseline, say baseline $k$, we collect two sets of data, one set is obtained through setting zero phase shift on $S$ and the other set is obtained through setting $\frac{\pi}{2}$ phase shift on $S$. 
In the first (second) set, we observe the event of detector $D_a$ silent and $D_b$ clicking for $m_k\left(\bar{m}_k\right)$ times in total detected events of one-detector-clicking $M_k\left(\bar{M}_k\right)$. We have the experimental data $\hat{\phi}_k$ of baseline $k$ as
\begin{equation}
    \hat{\phi}_k=\frac{1}{2}\left(\cos ^{-1} q_k+\sin ^{-1} \bar{q}_k\right) \bmod 2 \pi,
\end{equation}
where $q_k=1-\frac{2 m_k}{M_k}$ and $\bar{q}_k=\frac{2 \bar{m}_k}{\bar{M}_k}-1$.
The relation between the actual value $\mathring{\phi}_k$ and its observed value $\hat{\phi}_k$ is
\begin{equation}\label{equ:definition_of_ek}
    \hat{\phi}_k=\left(\mathring{\phi}_k+e_k\right) \bmod 2 \pi, \forall k,
\end{equation}
where $e_k \in(-\pi, \pi]$ is the observation error. We introduce a new variable
$$
\psi_k=\mathring{\phi}_k-\hat{\phi}_k+\pi, \forall k\in\{1,2\cdots K\}
$$
and its modulo
$$
\mathring{\psi}_k=\left(\mathring{\phi}_k-\hat{\phi}_k+\pi\right) \bmod 2 \pi, \forall k\in\{1,2\cdots K\}.
$$

\textbf{Theorem 1:} Under the condition
$$|e_{k+1}-s_k e_k|<\pi,$$
the iterative formula:
\begin{equation}\label{equ:iterative_formula}
\mathring{\psi}_{k}=\frac{\mathring{\psi}_{k+1}+\mathring{r}_{k+1}}{s_k}+\frac{s_k-2}{s_k}\pi
\end{equation}
always holds, where $\mathring{r}_{k+1}=(\hat{\phi}_{k+1}-s_k\hat{\phi}_{k}+\pi)\bmod 2\pi$.

\textbf{Proof of Theorem 1:} Observe the identity in modulo arithmetic:
$$
(x \bmod 2 \pi+y) \bmod 2 \pi=(x+y) \bmod 2 \pi .
$$
Taking $\mathring{\phi}_{k+1}=(s_k\mathring{\phi}_k)\bmod2\pi$ into the definition of the variable $\mathring{\psi}_{k+1}$, we obtain
\begin{equation}
    \begin{aligned}\label{equ:iterative_relation_unsolved}
    \mathring{\psi}_{k+1}&=((s_k\mathring{\phi}_{k})\bmod 2\pi-\hat{\phi}_{k+1}+\pi)\bmod 2\pi\\ &= (s_k\mathring{\phi}_k-s_k\hat{\phi}_k - (\hat{\phi}_{k+1}-s_k\hat{\phi}_k) + \pi)\bmod 2\pi\\
    &=(s_k\mathring{\psi}_k+ \mathring{r}_{k+1}-(s_k-2)\pi)\bmod 2\pi,
\end{aligned}
\end{equation}
Taking Eq.~\eqref{equ:definition_of_ek} into $\mathring{\psi}_{k+1}$, $s_k\mathring{\psi}_{k}$ and $\mathring{r}_{k+1}$ independently, and we obtain
\begin{equation}\label{equ:single_baseline_proof}
    \begin{aligned}
    \mathring{r}_{k+1}&=(e_{k+1}-s_k e_k + \pi)\bmod2\pi\\
\mathring{\psi}_{k+1}&=(\pi-e_{k+1})\bmod2\pi\\
s_k\mathring{\psi}_{k}&=s_k(\pi-e_{k})\bmod2\pi.
\end{aligned}
\end{equation}
Since $e_k\in(-\pi,\pi)$ and $|e_{k+1}-s_k e_k|<\pi$, $\mathring{r}_{k+1}=e_{k+1}-s_k e_k + \pi$, $\mathring{\psi}_{k+1}=\pi-e_{k+1}$. Thus:
$$
\begin{aligned}
s_k\mathring{\psi}_{k}=s_k(\pi-e_{k})=\mathring{\psi}_{k+1}+\mathring{r}_{k+1}+(s_k - 2)\pi.
\end{aligned}
$$
The proof is finished.

Using Eq.~\eqref{equ:iterative_formula} iteratively we have:
\begin{equation}\label{equ:formula_of_psi0}
    \mathring{\psi}_{1}=\sum_{k=2}^{K} \frac{\mathring{r}_{k}+(s_{k-1}-2)\pi}{\prod_{i=1}^{k-1} s_{i}} +\frac{\psi_{K}}{\prod_{i=1}^{K-1}s_{i}}.
\end{equation}
Taking $\mathring{\psi}_1 = (\mathring{\phi}_1 - \hat{\phi}_1 + \pi) \bmod 2\pi$
into Eq.~\eqref{equ:formula_of_psi0} and taking the approximation $\psi_{K}=\pi$, we obtain the highly accurate estimated value of $\phi_1$:
\begin{equation}\label{equ:formula_of_phi0}
    \tilde{\phi}_1 = \hat{\phi}_1 - \pi 
    + \sum_{k=2}^{K-1} \frac{\mathring{r}_{k} + (s_{k-1} - 2)\pi}{\prod_{i=1}^{k-1}s_{i}}
    + \frac{\pi}{\prod_{i=1}^{K-1}s_{i}},
\end{equation}
The precision of our method reaches $\frac{\pi}{\prod_{i=1}^{K-1}s_{i}}$ in the quantity of uncertainty, and the precision improves much faster than the increase of the number of baselines.
The estimated value of $\theta$ is:
\begin{equation}\label{equ:definition_theta}
    \tilde{\theta}=\frac{\tilde{\phi}_1 \lambda}{2\pi L_1}.
\end{equation}
Eq.~\eqref{equ:definition_theta} holds not only for our method but also for other existing schemes.

\subsection{Piecemeal method with reference star}\label{sec:reference_star_method}
Consider two remote objects: the reference star and the
target star. Suppose that the angles of the two remote objects are $\theta_R$ and $\theta_T$, respectively. Our goal is to detect the angle of the target star relative to
the reference star, whose value can be around the magnitude order of 1 arcsecond.
Denote the relative angle as
\begin{equation}\label{equ:Defi_gamma0}
    \Gamma=\Gamma_0 + \delta\theta
\end{equation}
where the value of $\Gamma_0$ is known to us and $\delta\theta$ is an
unknown small value whose range is $\delta\theta\in(0,\theta_0]$. The value of $\theta_0$ is known to us.
Consider the set-up imperfections that the baseline length and the angles of the objects fluctuate within a range. When the baseline length error and the orientation error of baseline $k$ are $\delta L_k$ and $\delta\theta_k$, the state of the single photon coming from the target star and is captured by the telescopes of baseline $k$ can be written as
\begin{equation}\label{equ:Sk_T}
    \begin{aligned}
|\mathcal{S}_k^T\rangle=&\frac{1}{\sqrt{2}}(|01\rangle+\mathrm{e}^{i\frac{2\pi(L_k+\delta L_k)}{\lambda}\sin{(\theta_T+\delta\theta_k)}}|10\rangle)\\
=&\frac{1}{\sqrt{2}}(|01\rangle+\mathrm{e}^{i\Phi_k^T}|10\rangle),
\end{aligned}
\end{equation}
where $\Phi_k^T=\phi_k^T+\delta\phi_k^T$ , $\phi_k^T=\frac{2\pi L_k \sin{(\theta_T)}}{\lambda}$, $\delta\phi_k^T=\frac{2\pi\delta L_k}{\lambda}\cos{\theta_T}+\frac{2\pi L_k}{\lambda}\sin{\delta\theta_k}\sin{\theta_T}$.
The state of single-photon coming from the reference star is
\begin{equation}\label{equ:Sk_R}
\begin{aligned}
|\mathcal{S}_k^R\rangle=&\frac{1}{\sqrt{2}}(|01\rangle+\mathrm{e}^{i\frac{2\pi(L_k+\delta L_k)}{\lambda}\sin{(\theta_R+\delta\theta_k)}}|10\rangle)\\
=&\frac{1}{\sqrt{2}}(|01\rangle+\mathrm{e}^{i\Phi_k^R}|10\rangle),
\end{aligned}
\end{equation}
where $\Phi_k^R=\phi_k^R+\delta\phi_k^R$, $\phi_k^R=\frac{2\pi L_k \sin{(\theta_R)}}{\lambda}$, $\delta\phi_k^R=\frac{2\pi\delta L_k}{\lambda}\cos{\theta_R}+\frac{2\pi L_k}{\lambda}\sin{\delta\theta_k}\sin{\theta_R}$.
$\delta\phi_k^T - \delta\phi_k^R$ is a second-order quantity compared to $\phi_k^T-\phi_k^R$. If we take 
\begin{equation}\label{equ:delta_Delta}
    \delta\phi_k^T - \delta\phi_k^R =\frac{2\pi\cos{\theta_R}\Gamma_0}{\lambda}\delta L_k - \frac{2\pi L_k\sin{\theta_R}\Gamma_0}{\lambda}\delta\theta_k = 0,
\end{equation}
we obtain
$$
\begin{aligned}
    \Phi_k^T-\Phi_k^R&=\phi_k^T-\phi_k^R\\
    &=\frac{2\pi L_k \sin{(\theta_R+\Gamma)}}{\lambda}-\frac{2\pi L_k \sin{\theta_R}}{\lambda}\\&\approx \frac{2\pi L_k}{\lambda}\cos{\theta_R}(\delta\theta\cos{\Gamma_0}+\sin{\Gamma_0}).
\end{aligned}
$$
Define
\begin{equation}
    \mathring{\Delta}_k=(\Phi_k^T-\Phi_k^R-\frac{2\pi L_k}{\lambda}\sin{\Gamma_0}\cos{\theta_R})\bmod 2\pi.
    \label{equ:definition_delta}
\end{equation}
The experimental value of baseline $k$ is $\hat{\Delta}_k$, and the error of the experimental value is $e_k$:
\begin{equation}\label{equ:11}
    \hat{\Delta}_k = (\mathring{\Delta}_k + e_k)\bmod 2\pi, \forall k\in\{1,2\cdots K\}.
\end{equation}
We introduce the new variable
$$\mathring{\Psi}_k=(\mathring{\Delta}_k-\hat{\Delta}_k+\pi)\bmod 2\pi.$$

\textbf{Theorem 2:}Under the condition
$$|e_{k+1}-s_k e_k|<\pi,$$
we have the iterative formula:
\begin{equation}\label{equ:relation_delta}
\mathring{\Psi}_k=\frac{\mathring{\Psi}_{k+1}+\mathring{\gamma}_{k+1}+(s_k-2)\pi}{s_k},
\end{equation}
where
$\mathring{\gamma}_{k+1}=(\hat{\Delta}_{k+1}-s_k\hat{\Delta}_k+\pi)\bmod 2\pi$.

The proof of \textbf{Theorem 2}  follows the same approach as that of \textbf{Theorem 1}, and therefore will not be elaborated here.

Eq.~\eqref{equ:relation_delta} leads to the conclusion:
\begin{equation}\label{equ:result_delta}
    \tilde{\Delta}_1=\hat{\Delta}_1 - \pi + \sum_{k=2}^{K}(\frac{\mathring{\gamma}_k+(s_{k-1}-2)\pi}{\prod_{i=1}^{k-1}s_i})+\frac{\pi}{\prod_{i=1}^{K-1}s_i}.
\end{equation}
The RHS of Eq.~\eqref{equ:result_delta} contains only experimental data of different baselines with error, but the difference between the estimated value $\tilde{\Delta}_1$ and the true value $\mathring{\Delta}_1$ will be smaller than $\frac{\pi}{\prod_{i=0}^{K-1}s_i}$.
Finally, we get the estimated value of $\Gamma$:
\begin{equation}\label{equ:result_TOR}
    \tilde{\Gamma}=\frac{\tilde{\Delta}_1}{\cos{\theta_R}}\frac{\lambda}{2\pi L_1}+\Gamma_0.
\end{equation}
Here we have ignored the second order term $\delta\phi_k^T - \delta\phi_k^R$. Below, we consider the effect of this term in detail.




\section{Performance with set-up imperfections}
In practical experiments, setup imperfections may arise, such as errors in baseline lengths and orientations, which significantly impact angular measurement precision. Our piecemeal approach offers an advantage in mitigating these noise effects. In this section, we demonstrate the robustness of the extended piecemeal method against baseline length and orientation errors under different baseline numbers $K$ and baseline length ratios $s_k$.

We shall use the average failure probability and the uncertainty of the observed angle results to judge the performance of different methods. The failure probability is defined as:
\begin{equation}\label{equ:failure_probability}
    \langle\epsilon\rangle=\overline{P\left(\left|\tilde{\phi}_1-\mathring{\phi}_1\right|>\frac{\pi}{2^K}\right)}
\end{equation}
for a set of $\{\mathring{\phi}_1\}$, where each element is randomly chosen between $[0, 2\pi)$. An experimental result is marked as successful if the output of the method, $\tilde{\phi}_1$, correctly provides the first $K$ bits of $\mathring{\phi}_1$ in binary form. In our study here, we shall generate the initial data to simulate the experimental data and repeat the simulation by $n$ times for one point. Among these $n$ calculation results, the ratio $\frac{n-m}{n}$ is the failure probability where $m$ is the number of successful results.


\subsection{Robustness of our method with baseline length error}
Earlier, we simply take $\delta\phi_k^T - \delta\phi_k^R  = 0$ in Eq.~\eqref{equ:delta_Delta}. However, in practice, its value cannot be zero due to the set-up imperfections, such as the baseline length error and the orientation error of different baselines.
There are errors in the value of \(\hat{\Delta}_k=(\Phi_k^T-\Phi_k^R-\frac{2\pi L_k}{\lambda}\sin{\Gamma_0}\cos{\theta_R})\bmod 2\pi\) defined in Eq.~\eqref{equ:definition_delta} if we simply take $\delta\phi_k^T - \delta\phi_k^R =0$.

Suppose we have assumed the length of baseline $k$ is $L_k$. Due to the baseline length error $\delta L_k$, the actual baseline length is $L_k + \delta L_k$. The baseline length error $\delta L_k$ and the orientation error $\delta\theta_k$ cause error $e_k$ in relative phase $\Delta_k$, explicitly 
\begin{equation}
    \begin{aligned}
        e_{k} &= \delta\phi_k^T - \delta\phi_k^R\\
        &=\frac{2\pi\cos{\theta_R}\Gamma}{\lambda}\delta L_k - \frac{2\pi L_k\sin{\theta_R}\Gamma}{\lambda}\delta\theta_k
    \end{aligned}
\end{equation}
and $\hat{\Delta}_k=(\Delta_k+e_k)\bmod 2\pi$.
According to \textbf{Theorem 2}, our method works well as long as $|e_{k+1} - s_k e_{k}|<\pi$ for all baselines. More strictly, we have the following sufficient condition for Eq.~\eqref{equ:result_TOR}:

\textbf{Fact 1}: Disregarding baseline length error $\delta L_k$ or orientation error $\delta\theta_k$, Eq.~\eqref{equ:result_TOR} of our piecemeal method gives the correct result provided that
$$
|e_{k}| = |\delta\phi_k^T - \delta\phi_k^R| < \frac{\pi}{s_k + 1}
$$
holds for all baselines.

However, given baseline length error $\delta L_k$ no matter how small it is, it leads to error at least $\delta\theta=\frac{\delta L_k}{L}\Gamma\cos{\theta_R}$ in existing single-baseline methods with baseline length $L$. Here we have assumed no orientation error.

\begin{figure}
    \centering
    \begin{minipage}[t]{0.98\linewidth}
        \centering
        \includegraphics[width=\textwidth]{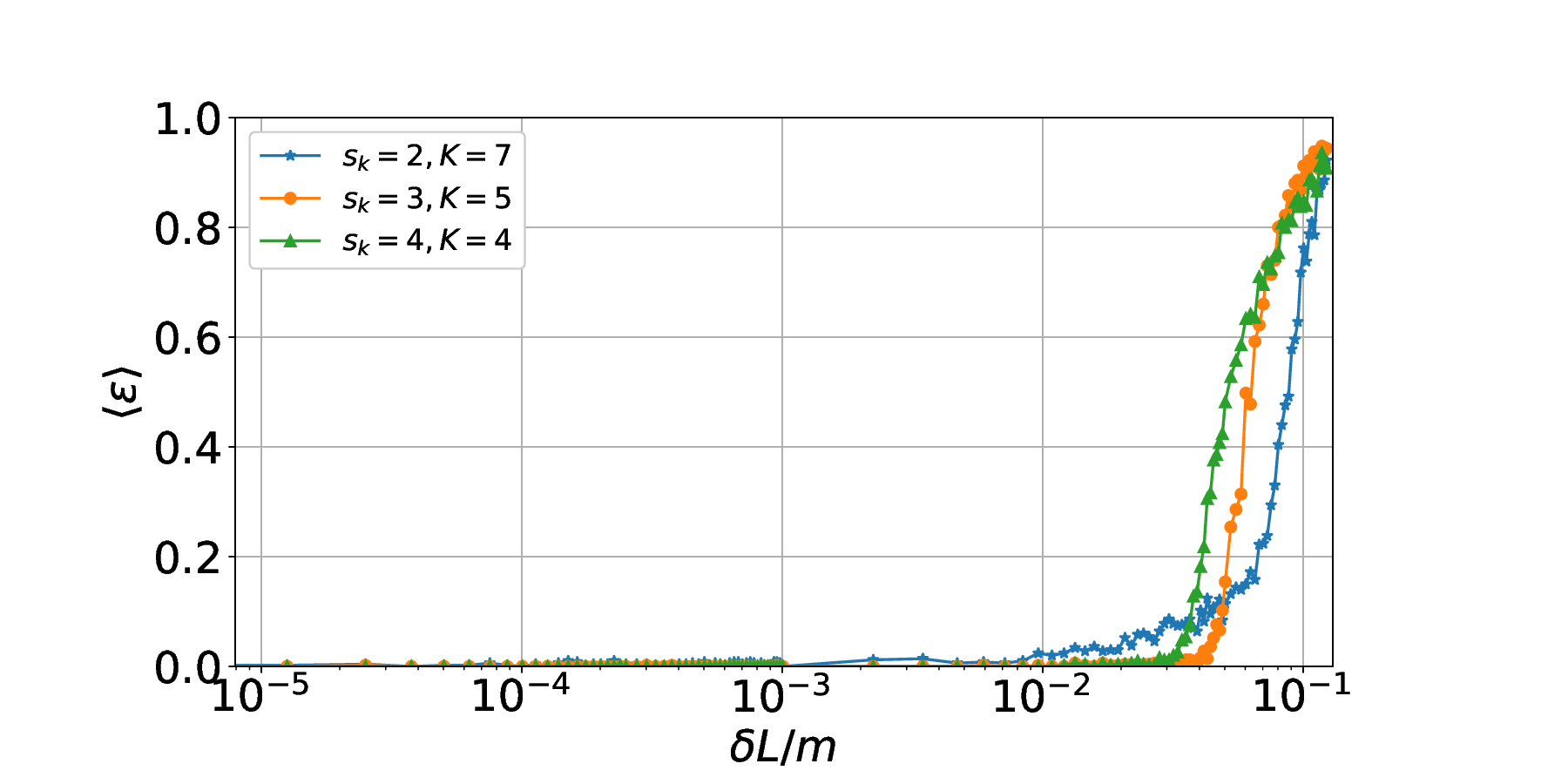}
        \caption*{(a)}
    \end{minipage}
    \hfill
    \begin{minipage}[t]{0.98\linewidth}
        \centering
        \includegraphics[width=\textwidth]{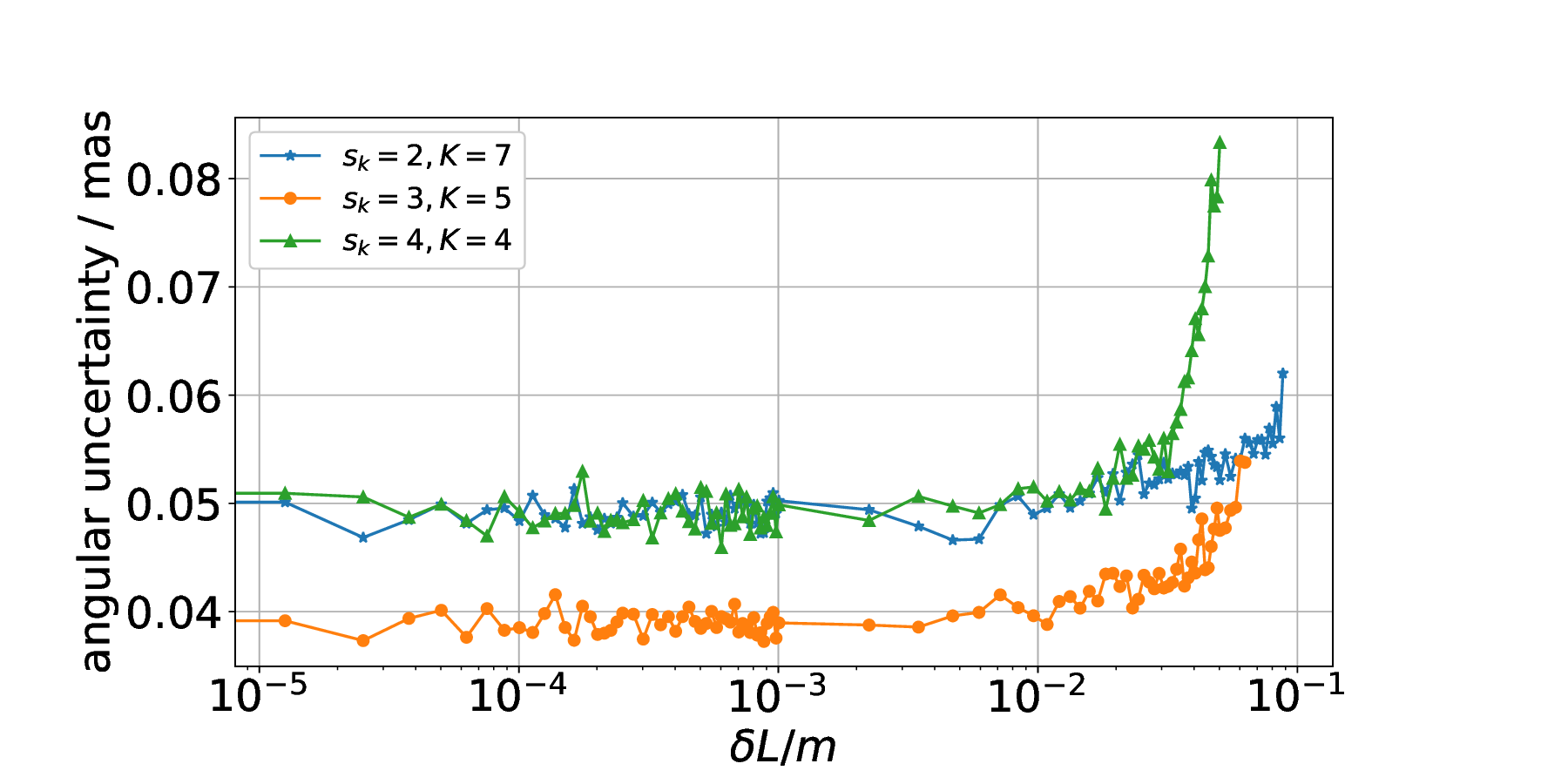}
        \caption*{(b)}
    \end{minipage}
    
    \caption{The variation of the average failure probability and angular uncertainty with the maximum baseline length error in our generalized piecemeal scheme under different \( (s_k, K) \) setups. For each point, we repeat the calculation by hundreds of times, with the input data pairs ($\theta_T=\theta_T^0+\delta\theta_T, \theta_R=\theta_R^0+\delta\theta_R$), where $\delta\theta_T$ is randomly chosen from the range \(\delta\theta_T \leq 5\) mas and $\delta\theta_R$ is randomly chosen from the range \(\delta\theta_R \leq 5\) mas. In each calculation, each setup used \( 2 \times 10^3 \) photons. We then calculate the failure probability based on the statistics of number of successful results and number of failure results among those hundreds of calculations. (a) Comparison of average failure probabilities defined in Eq.~\eqref{equ:failure_probability} with baseline length errors. (b) Comparison of resolution of different setups with baseline length errors. The vertical axis is the angle uncertainty.}
    \label{fig:comparahension_Ksk}
\end{figure}

We present in Fig.\ref{fig:comparahension_Ksk} the relationship between the average failure probability, angular uncertainty, and the maximum baseline length error for different values of \( s_k \) and different numbers of baselines \( K \) as specified in the \textbf{notations}. All three datasets are simulated using \( 2 \times 10^3 \) photons. The shortest baseline length in each setup is identical (\( L_1 = 10m \)), and the maximum baseline lengths are the same for the cases of \( (s_k=2, K=7) \) setup and \( (s_k=4, K=4) \) setup (\( L_K = 640m \)), while the maximum baseline length for \( (s_k=3, K=5) \) setup is longer (\( L_K = 810m \)). 
From Fig.\ref{fig:comparahension_Ksk}(a), we observe that reducing the number of baselines while increasing the baseline length ratio does not significantly decrease the maximum tolerable baseline length error of our method. Meanwhile, as shown in Fig.\ref{fig:comparahension_Ksk}(b), within the tolerable range of baseline length errors, our approach consistently achieves an angular measurement accuracy of 0.05 mas.

\begin{figure}
    \centering
    \begin{minipage}[t]{0.98\linewidth}
        \centering
        \includegraphics[width=\textwidth]{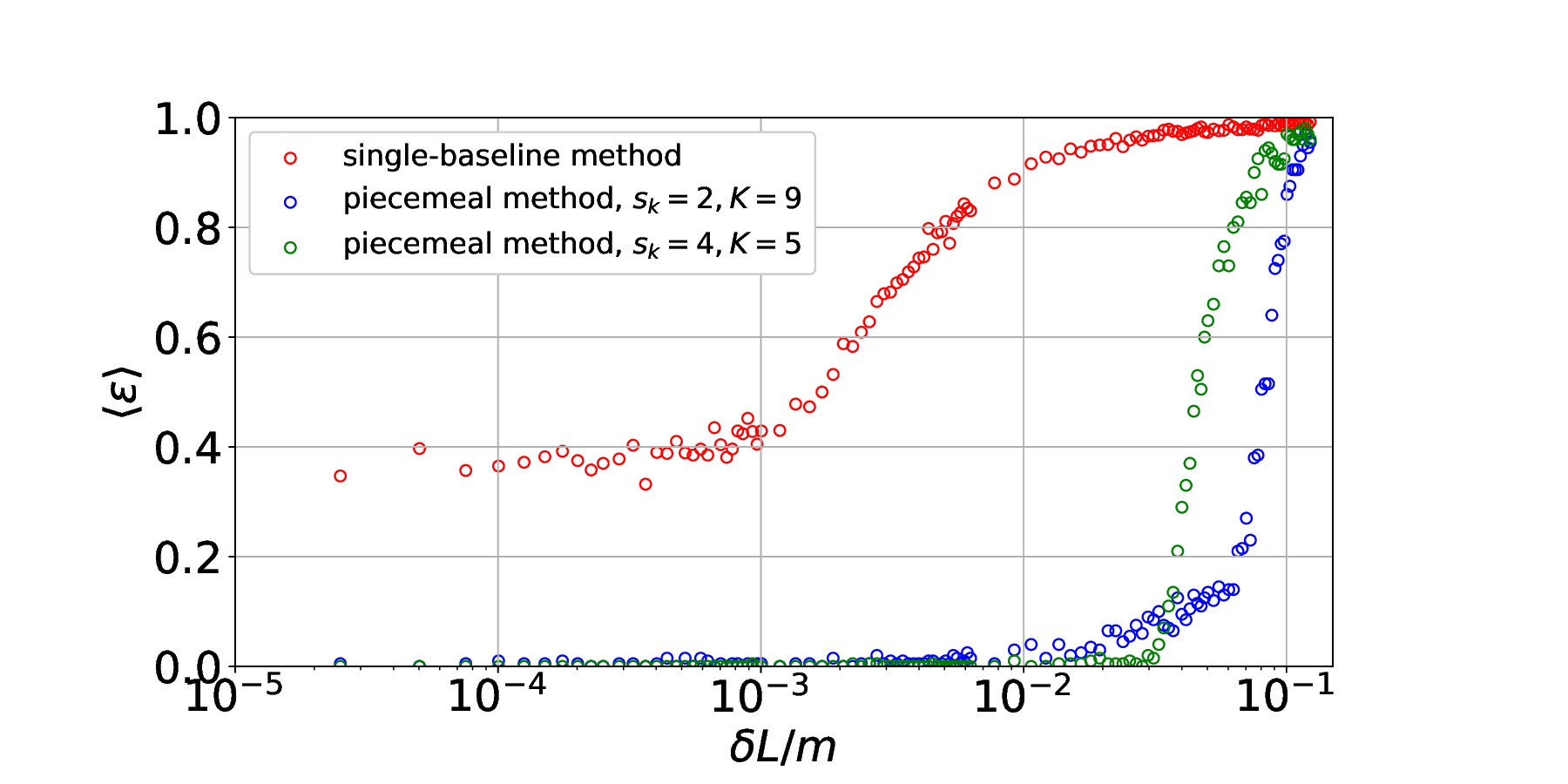}
        \caption*{(a)}
    \end{minipage}
    \hfill
    \begin{minipage}[t]{0.98\linewidth}
        \centering
        \includegraphics[width=\textwidth]{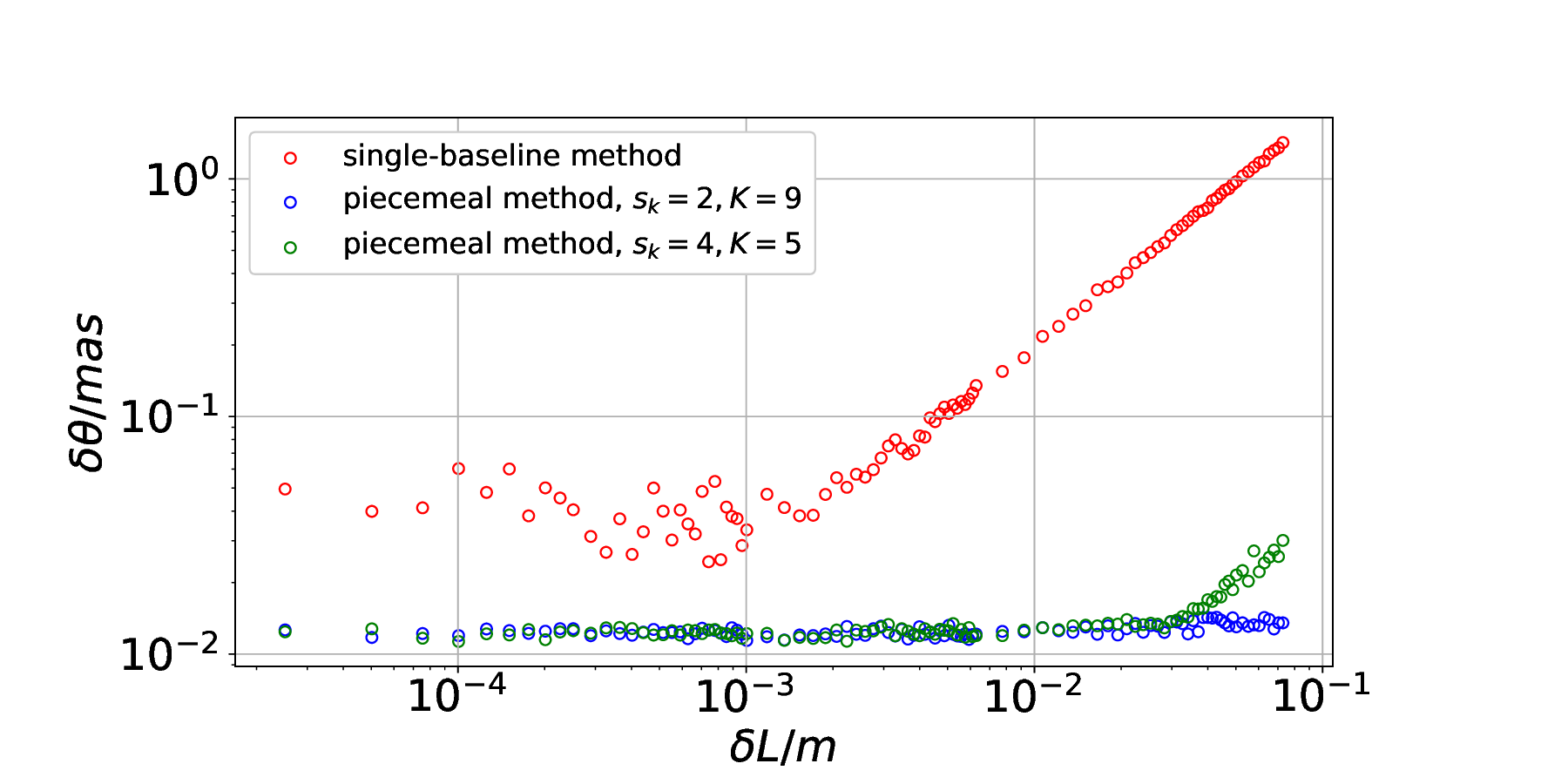}
        \caption*{(b)}
    \end{minipage}
    
    \caption{Performance comparison of our piecemeal method and existing single-baseline method. In the numerical simulation, we have considered the effects of baseline length random errors. For each point, we repeat the calculation by hundreds of times, with the input data pairs ($\theta_T=\theta_T^0+\delta\theta_T, \theta_R=\theta_R^0+\delta\theta_R$), where $\delta\theta_T$ is randomly chosen from the range \(\delta\theta_T \leq 5\) mas and $\delta\theta_R$ is randomly chosen from the range \(\delta\theta_R \leq 5\) mas. In each calculation, both methods used \( 2 \times 10^3 \) photons. We then calculate the failure probability based on the statistics of number of successful results and number of failure results among those hundreds of calculations. (a) Comparison of average failure probabilities defined in Eq.~\eqref{equ:failure_probability} with baseline length errors. The blue line is the result of \cite{huang_imaging_2022}, and the red line is the result of our piecemeal method\cite{leng_piecemeal_2024}. (b) Comparison of resolution of different methods with baseline length errors. The vertical axis is the uncertainty of observation. The blue line is the result of \cite{huang_imaging_2022}, and the red line is the result of our piecemeal method\cite{leng_piecemeal_2024}.}
    \label{fig:error_against_Lfluc}
\end{figure}

To compare the robustness of our scheme and existing single-baseline scheme, we take $\Gamma_0 = 500 mas$ in Eq.~\eqref{equ:Defi_gamma0}. 
The numerical simulation results are presented in Fig.\ref{fig:error_against_Lfluc}. Choosing the wavelength of incident light to be $\lambda = 550nm$.
From these numerical simulations, it is evident that our method holds a distinct advantage in the presence of baseline errors, and this advantage becomes increasingly pronounced when the photon count is low.
As shown in Fig.\ref{fig:error_against_Lfluc}(a), our numerical simulation shows correct results of our piecemeal method with baseline length error up to $\delta L_k=0.08m$ for all $\{k\}$ when choosing  \(s_k = 2\) in item 1 of \textbf{notations} presented earlier, or up to $\delta L_k=0.05m$ for when choosing  \(s_k = 4\). Under the same failure probability condition, the single-baseline method \cite{huang_imaging_2022} can only tolerate a baseline length error smaller than $1mm$.
Fig.\ref{fig:error_against_Lfluc}(b) shows the uncertainty of observed angles with baseline length errors for different methods. In particular, when baseline length error is around $10mm$, the single-baseline method has an uncertainty of about $0.2mas$, which is in agreement with the existing experiment \cite{ten_brummelaar_first_2005}. While with the same errors, our method has an uncertainty of about $10 \mu as$, which shows a precision improvement of about 20 times compared with the existing single-baseline method.

\section{Conclusion}
In conclusion, we for the first time took into account the influence of practical conditions on weak-light interference schemes and demonstrated the robustness of our piecemeal method under set-up imperfections. The accuracy of our method is better than existing single-baseline weak-light-interference schemes \cite{gottesman_longer-baseline_2012, khabiboulline_optical_2019, huang_imaging_2022, marchese_large_2023}. Our method can be used to enhance the angular resolution of existing telescope systems over faint targets, providing more accurate data for critical observational studies in general relativity and cosmology.

We acknowledge the financial support in part by National Natural Science Foundation of China grant No.12174215 and No.12374473. This study is also supported by the Taishan Scholars Program. We thank Prof. Y Cao of USTC for useful discussions.

\nocite{*}
\bibliography{ref}

\end{document}